\documentclass[twocolumn,epsfig,showpacs,oneside,amsmath,amssymb,floatfix]{revtex4}

\usepackage{epsfig}

\begin{document}

\title{Zn-impurity effect and interplay of $s_{\pm}$-
and $s_{++}$-pairings in Fe-based superconductors}

\author{Zi-Jian Yao$^1$, Wei-Qiang Chen$^{2,1}$, Yu-ke Li$^3$,
  Guang-han Cao$^4$, Hong-Min Jiang$^{1,3}$, Qian-En Wang$^1$, Zhu-an
  Xu$^4$, Fu-Chun Zhang$^{1,4}$}
\affiliation{$^1$ Department of Physics and Center of Theoretical and Computational Physics, The University of Hong Kong, Hong Kong, China\\
  $^2$ Department of Physics,South University of Science and Technology of China, Shenzhen, China\\
  $^3$ Department of Physics,Hangzhou Normal University, Hangzhou, China\\
  $^4$ Department of Physics,Zhejiang University, Hangzhou, China\\
}

\date{\today}
\begin{abstract}
  We report theoretical and experimental studies of the effect of
  Zn-impurity in Fe-based superconductors. Zn-impurity is expected to
  severely suppress sign reversed s$_\pm$ wave pairing. The
  experimentally observed suppression of T$_c$ under Zn-doping strongly depends on the materials and the charge
  carrier contents, which suggests competition of  $s_{++}$ and $s_{\pm}$ pairings
  in Fe-base superconductors.  We study a model incorporating both $s_{++}$ and $s_{\pm}$ pairing couplings by using Bogoliubov de-Gennes equation, and show that the Zn-impurity strongly suppresses $s_{\pm}$
  pairing and may induce a transition from $s_{\pm}$ to
  $s_{++}$-wave. Our theory is consistent with various experiments on
  the impurity effect. We present new experimental data on the
  Zn-doping SmFe$_{1-x}$Zn$_x$AsO$_{0.9}$F$_{0.1}$ of T$_c=$ 50K, in further support of our proposal.
\end{abstract}

\maketitle

\section{Introduction}
One of the most important issues in the high $T_c$ Fe-based
superconductors (FeSC) is their pairing symmetry
~\cite{ding:47001,sato:047002,liu:177005}. Theories based on
antiferromagnetic (AF) spin fluctuations have predicted $s_{\pm}$
pairing, where the superconducting (SC) order parameters on the hole
and electron Fermi pockets have opposite
signs~\cite{mazin:057003,wang:047005}.  The proposed symmetry is
consistent with a number of experiments, such as the spin resonance
peak in neutron scattering~\cite{christianson:930}, sensitive SC
junction data~\cite{chen:260,teague:087004}, and quasiparticle
interference in tunneling experiments~\cite{hanaguri:23042010}.
However, the pairing symmetry in FeSC may not be universal, and there
are evidences for different pairing structures as discussed in a
recent review \cite{hirschfeld:124508}.

The effect of disorder to the superconductivity is an important test
to the pairing symmetry.  According to Anderson's theorem, the
conventional s-wave superconductivity is insensitive to non-magnetic
impurities.  The sign reversed $s_{\pm}$ superconductivity is,
however, sensitive to non-magnetic impurities which scatter inter-band
electrons. Replacement of part of Fe-atoms by Co or Ni in a parent
compound of FeSC leads to superconductivity.  However, the role of the
Co or Ni doping is more subtle and remains controversial.  One
scenario is that the doping introduces additional electron carriers. This scenario is supported by the
angle resolved photoemission spectroscopy, which indicates the shrinking of
the hole pockets~\cite{kaminski:419}.  On the other hand, recent resonant
photoemission spectroscopy and density functional calculations
indicate that Co doping is covalent and introduces
disorder~\cite{sawatzky:077001}. It is plausible that the Co doping introduces
both carriers and disorder~\cite{ku:207003}.  Zn-ion has a $3d^{10}$
configuration, hence a very high electric potential to charge
carriers. Replacing a Fe-atom by Zn in FeSC introduces inter-band
scattering and is expected to severely suppress the $s_{\pm}$
superconductivity. Therefore, the Zn-doping is an effective test to
the $s_{\pm}$ pairing in FeSC.  There have been several experiments on
the Zn-doping effect on FeSC, including so-called 1111 compounds
LaFe$_{1-x}$Zn$_x$AsO$_{1-y}$F$_y$~\cite{li:053008,li:083008}, and more
recently 122 compounds
BaFe$_{2(1-x-y)}$Zn$_{2x}$Co$_{2y}$As$_2$ and SrFe$_{1.8-2x}$Zn$_{2x}$Co$_{0.2}$As$_2$~\cite{li:671}.
The results are
mixed at present, which appears to be strongly dependent of material and charge carrier content.  The
experimental data on the 1111 compounds may be divided into two
categories. The optimally doped LaFeAsO$_{0.9}$F$_{0.1}$
\cite{li:053008} is insensitive, but the over-doped
LaFeAsO$_{0.85}$F$_{0.15}$ is very sensitive to the
Zn-impurities~\cite{li:083008}. The effect of Zn-doping on Co-doped
122 compounds clearly shows the suppression of superconducting
transition temperature $T_c$, but the reduction is much slower than
the theory predicted~\cite{li:671}. A careful examination indicates
that the suppression of $T_c$ may be saturating at large Zn-doping to
some of the compounds.  Note that it is not easy to dope Zn into the Fe lattices
uniformly even under high pressure, and reliable data is only available up to
$6\%$ Zn-doping at present. Therefore the experimental data are not complete.
Nevertheless, the available experiments on Zn-doping indicate
complexity of the effect, and suggest possible competition of sign
changed $s_{\pm}$ and sign unchanged $s_{++}$ pairings in FeSC.

In this paper, we use a two-orbital model for FeSC including both
on-site (or $s_{++}$) pairing coupling $g_0$ and next nearest neighbor
(NNN) intersite (or $s_{\pm}$) pairing coupling $g_2$ to study
Zn-impurity effect, which may help understand the complex result of
the Zn doping effect on 1111 and 122 compounds. We apply Bogliubov
de-Gennes (BdG) equation to study the model on a finite-size
system. The two SC pairings in the multi-band system show interesting
interplay. They may mix but also compete with each other. The disorder
strongly suppresses the intersite pairing, and its effect to the
superconductivity depends on the strength of $g_0$.  For large $g_0$,
$g_2$ plays little role and the pairing is $s_{++}$ and is robust
against the disorder. For small $g_0$, the pairing is $s_{\pm}$ and
the disorder strongly suppresses superconductivity. For moderate value
of $g_0$, the disorder may enhance the on-site pairing and induce a
transition from $s_{\pm}$ to $s_{++}$ superconductivity.  We further
study the interplay between $g_0$ and $g_2$ in a clean system and show
that the disorder effect on the gap functions is similar to the
reduction of $g_2$.  Our theory is consistent with the Zn-doped
impurity experiments on 1111 and 122 compounds, and suggests multi-pairing
couplings in some of the FeSC.  We present our new experimental data
of the Zn-impurity effect on the very high $T_c=50$K Sm-1111
compound. The lattice constant measurement show that the Zn-atoms are
doped into the Fe-lattice uniformly up to $6\%$. The results appear to
indicate possible saturation of $T_c$ under the Zn doping, consistent
with the present theory.

\section{Model Hamiltonian and Mean Field Theory}
We consider a model Hamiltonian
\begin{equation}
  H=H_0 + H_{\mathrm{pair}} + H_{\mathrm{imp}},
\end{equation}
which includes a tight-binding kinetic term $H_0$, a pairing
interaction $H_\mathrm{pair}$, and a disordered term $H_{\mathrm{imp}}$. For $H_0$,
we consider a two-orbital model with $d_{xz}$ (orbital
1) and $d_{yz}$ (orbital 2) as proposed by Raghu et
al.~\cite{raghu:220503}.
\begin{equation}
  H_0=\sum_{\left < i{\alpha},j{\beta}\right >\sigma}
  C_{i\sigma}^{\dagger}\hat{h}_{ij}C_{j\sigma},
\end{equation}
where $C_{i\sigma}^{\dagger}= (c_{i1,\sigma}^{\dagger},
c_{i2,\sigma}^{\dagger})$, and $\hat{h}_{ij}^{\alpha, \beta}=
t_{i,j}^{\alpha\beta}$ is the electron hopping term between orbital
$\alpha$ at site $i$ and orbital $\beta$ at site $j$ on a
2-dim. square lattice of Fe-atoms (lattice constant $a=1$). While this
model may be an over-simplified one to describe many detailed material
properties of FeSC, it should capture the basic feature of the
disorder effect to the pairing in systems with multi-Fermi
surfaces. The non-vanishing hopping matrix elements are
$t_{i,i}^{\alpha \alpha} = -\mu$, $t_{i,i+\hat{y}}^{11} = t_2$,
$t_{i,i+\hat{y}}^{22} = t_1$, $t_{i,i+\hat{x}+\hat{y}}^{\alpha \alpha}
= t_3$, $t_{i,i+\hat{x}+\hat{y}}^{12} = t_{i,i+\hat{x}+\hat{y}}^{21}
=-t_4$. We choose $t_1=1$ as the energy unit, and $\mu=1.6$,
$t_2=-1.3$, $t_3=t_4=0.85$, which gives Fermi surfaces with hole
pockets near the $\Gamma$- and $M$ points, and electron pockets near
the $X$- and $Y$ points in an extended Brillouin zone as plotted in
Fig. 1.

\begin{figure}[htbp]
  \centering \includegraphics[width=0.25\textwidth]{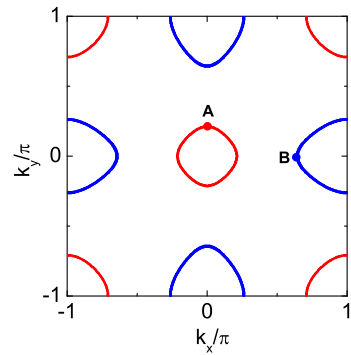}
  \caption{(Color online) Hole (red) and electron (blue) Fermi pockets obtained in
    the two-orbital model Eq. (2).  Points A and B are the
    representative $\vec k$ points for the hole and electron pockets,
    respectively.}
\end{figure}

We consider randomly distributed impurities on the lattice and
introduce an on-site repulsive potential on the Zn-impurity site,
\begin{equation}
  H_{\mathrm{imp}}= I\sum_{i \in{\mathrm{imp}}} \sum_ {\sigma}C_{i
    \sigma}^{\dagger}C_{i\sigma},
\end{equation}
where $i$ sums over all the impurity sites, and we
consider the large $I$ case ($I=24t_1$ in the actual calculation) to model the large repulsion to an electron at the Zn site. The
pairing Hamiltonian is modeled by
\begin{equation}
  \label{eq:4} H_\mathrm{pair} =
  -\sum_{<ij>}(V_{ij}c^{\dagger}_{i\alpha\uparrow}c^{\dagger}_{j\beta\downarrow}c_{j
    \beta\downarrow} c_{i \alpha \uparrow}+h.c.),
\end{equation}
where the pairing coupling $V_{ij}$ includes an on-site term
$g_0>0$ and an NNN intersite term $g_2$,
\begin{equation}
  V_{ij}=g_0\delta_{i,j}+
  g_2\sum_{\vec{\tau}}\delta_{j,i+\vec{\tau}}. \label{realv}
\end{equation}
with $\vec{\tau}$ the vector of the two NNN site displacement.
Note that $g_0$ term favors $s_{++}$ and $g_2$ term
favors $s_{\pm}$ symmetry.

We introduce a mean field gap function $\Delta_{ij}^{\alpha
  \alpha}=V_{ij} \left \langle c_{j \alpha \downarrow} c_{i \alpha
    \uparrow} \right \rangle$.  Our calculations show that the
inter-orbital pairing $\Delta_{ij}^{12}$ is very tiny and will be
neglected below.  The BdG equation for the mean field Hamiltonian then
reads
\begin{eqnarray}
  \label{bdg}
  \sum_j \left(
  \begin{array}{cc}
    \hat{h}_{ij} & \hat{\Delta}_{ij} \\
    \hat{\Delta}^{\ast}_{ij} & -\hat{h}^{\ast}_{ij,\bar{\sigma}}
  \end{array}
  \right)
  \left(
  \begin{array}{c}
    \mathbf{u}^{n}_{j,\sigma} \\
    \mathbf{v}^{n}_{j,\bar{\sigma}} \\
  \end{array}
  \right)= E_{n}\left(
  \begin{array}{c}
    \mathbf{u}^{n}_{i,\sigma} \\
    \mathbf{v}^{n}_{i,\bar{\sigma}} \\
  \end{array}
  \right),
\end{eqnarray}
with $\hat{\Delta}_{ij}=\Delta_{ij} \hat{I}$, and $\hat{I}$ an
identity matrix.
\begin{small}$\mathbf{u}_{i,\sigma}=\left(\begin{array}{c}
    u_{i1,\sigma} \\ u_{i2,\sigma}\end{array}\right)$\end{small}. The
self-consistent equation for the gap function is
\begin{eqnarray}
  \label{sc}
  \Delta_{ij}^{\alpha \alpha}&=&\frac{V_{ij}}{4}\sum_{n}(u^{n}_{i\alpha,\sigma}
  v^{n\ast}_{j\alpha,\bar{\sigma}}
  +v^{n\ast}_{i\alpha,\bar{\sigma}}u^{n}_{j\alpha,\sigma})\times \nonumber\\
  &&\tanh(\frac{E_{n}}{2k_{B}T})
\end{eqnarray}
For the form of $V_{ij}$ in Eq.~\eqref{realv}, we define
$\Delta_{0}^{\alpha \alpha}(i)= \Delta_{ii}^{\alpha \alpha}$, and $\Delta_{2}^{\alpha \alpha}(i)= \sum_{\vec \tau}
\Delta_{i,i+\vec \tau}^{\alpha \alpha}/4$.

\section{Numerical Results}
We now discuss the numerical solutions of $H$. In our calculations,
for each impurity content, the impurity positions are randomly
distributed and the statistical averages are taken over 400 times.  We
consider three typical cases: (i) $g_0$ is large and dominant; (ii)
$g_2$ is large and $g_0$ is weak; and (iii) $g_2$ is dominant but $g_0$ is moderately
large. In case (i), the SC pairing is always $s_{++}$ and the
superconductivity is robust against the impurity as we expect from the
Anderson theorem.

In Fig. 2 (a) and (b), we show the spatially averaged on-site and NNN
inter-site pairing amplitudes $\Delta_0=\Delta_0^{\alpha\alpha}$ and
$\Delta_2=\Delta_2^{\alpha\alpha}$ as functions of the impurity
concentration $n_{\mathrm{imp}}$ for cases (iii) and (ii). Also shown
are the gaps at the hole pocket (point A [$(0,0.22\pi)$]) and at the
electron pocket (point B [$(0.62\pi,0)$]), which are the Fourier
transform of the impurity averaged gaps in real space. In the case
(ii) of weak on-site pairing, the impurities strongly suppress
$\Delta_2$ as shown in the Fig. 2(b). $\Delta_0$ is tiny and the SC
gap functions $\Delta_A$ and $\Delta_B$ monotonically decrease as
$n_{\mathrm{imp}}$ increases. Because of the finite lattice size, our
study is limited to the short coherence length or the strong pairing
coupling cases, which require $n_{\mathrm{imp}} \approx 0.15$ to
destroy the superconductivity.  We expect this value to be much
smaller in weaker pairing coupling cases.

Case (iii) is most interesting, and our theory shows an impurity
driven phase transition from $s_{\pm}$ to $s_{++}$ pairings. In the
absence of impurity, $g_2$ dominates and the pairing is $s_{\pm}$.  As
shown in Fig. 2(a), the pairing symmetry remains to be $s_{\pm}$ at
$n_{\mathrm{imp}} < 0.02$, and the gap amplitudes on $k$ points A and
B are monotonically suppressed as $n_{\mathrm{imp}}$ increases. At $
0.02 < n_{\mathrm{imp}} < 0.05$, $\left| \Delta_2 \right|$ decreases, and
$\left| \Delta_0 \right|$ increases.  At $n_{\mathrm{imp}}> 0.05$, both $\Delta_A$
and $\Delta_B$ are positive and we have $s_{++}$ pairing.  It is
interesting to note that the on-site pairing may be enhanced by the
impurities due to the suppression of the NNN pairing.

\begin{figure}[htbp]
  \center
  \epsfig{figure=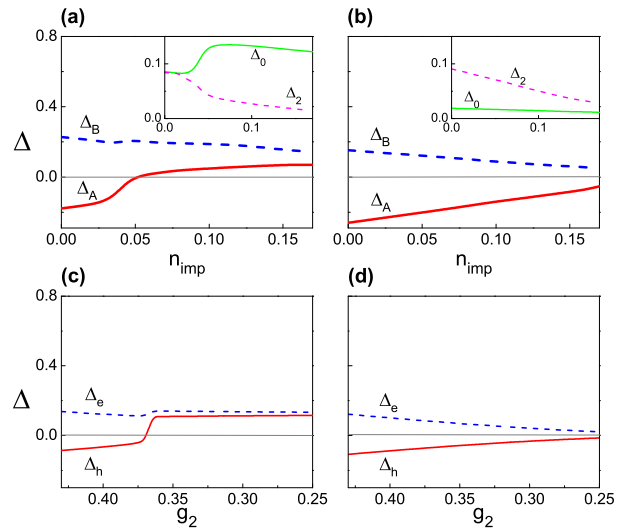,width=0.45\textwidth}
  \caption{(Color online) Upper panel: The gap functions at hole and
    electron Fermi pockets $\Delta_h$ and $\Delta_e$ as functions of
    impurity density $n_{\mathrm{imp}}$, obtained in the mean field
    solution for $H$. (a): for modestly strong on-site pairing
    coupling $g_0=1.8$; (b): for weak on-site coupling
    $g_0=0.8$. Insets: spatially averaged gap functions $\Delta_0$
    (on-site) and $\Delta_2$ (NNN inter site). In both cases, the NNN
    coupling $g_2=1.6$. Lower panel: SC gaps calculated by the
    simplified BCS formalism, with $N_e(0) = 0.12$, $N_h(0) = 0.1$,
    and $\omega_D = 0.8$. (c): $g_0=1.8$; (d): $g_0=0.8$.}
\end{figure}

In the case of weak on-site pairing, as shown in Fig. 2(b), the impurities
strongly suppresses $\Delta_2$. We have examined the
SC order parameters in real space and found that the disorder does not
result in severe pair-breaking effect to the on-site pairing measured
by $\Delta^{\alpha \alpha}_{i,i}$, whose peak amplitude is almost unaltered by the
impurities. On the other hand, the non-magnetic impurities not only
destroy NNN SC pairing order parameter $\Delta^{\alpha \alpha}_{i,i+\hat{x}+\hat{y}}$
in larger spatial areas, but also weaken the peak amplitude of the SC
pairing immensely.  

As we have demonstrated, the impurities suppress
the NNN pairing order parameter $\Delta_2$. This effect is similar to
the reduction of $g_2$ in the clean sample.  Therefore, tuning
$n_{\mathrm{imp}}$ in the disorder system is similar to tuning $g_2$
in a clean system~\cite{tuning_note}. Below we shall study SC order
parameters and $T_c$ in the model Hamiltonian $H$ as functions of
$g_2$ in the absence of disorder to mimic the impurity effect.  This
enables us to further reveal the interplay between the SC pairings of
$s_{++}$ and $s_{\pm}$. 


\section{$T_c$ Reduction: Theory and Experiments}
For a clean system, we have lattice translational symmetry, and the
gap function Eq. (7) becomes
\begin{equation}
  \label{gapeq}
  \Delta_m(\mathbf{k}) = - \sum_{\mathbf{k}^{\prime}}
  V_{mn}(\mathbf{k}, \mathbf{k}^{\prime})\frac{\tanh(\beta
    E_{n\mathbf{k}^{\prime}}/2)}{2E_{n\mathbf{k}^{\prime}}}\Delta_n(\mathbf{k}^{\prime})
\end{equation}
where $m, n$ are the band indices, $E_{n\mathbf{k}} =
\sqrt{\Delta_n({\mathbf{k}})^2 + \epsilon_n(\mathbf{k})^2}$,
$\epsilon_n(\mathbf{k})$ is the single particle energy.  The summation
is taken only in the vicinity of Fermi pockets with an energy cut-off
$\omega_D$.  The pairing potential $V_{m, n}(\mathbf{k}, \mathbf{k}')$
describes the coupling between gap function on various Fermi pockets,
and with Eq.~\eqref{eq:4}, we have
\begin{align}
  \label{eq:1}
  V_{mn}(\mathbf{k}, \mathbf{k}') &= \sum_{\alpha} U_{m \alpha}(-\mathbf{k}) U_{m \alpha}(\mathbf{k}) U_{n
    \alpha}(\mathbf{k}') U_{n \alpha} (-\mathbf{k}')
  \nonumber\\
  & \phantom{=} \times \left( g_0 + 4 g_2 \cos q_x \cos q_y \right),
\end{align}
where $m,n$ are band indices, $\alpha$ is orbital index,
$U(\mathbf{k})$ is the transformation matrix between bands and
orbitals. 

In our two-orbital model, there are four Fermi pockets, two for hole
bands at $\Gamma$ and M points respectively and two for electron bands
at X and Y points respectively, which makes it very difficult to solve
Eq.~\eqref{eq:1} analytically.  So in the following, we will ignore
the size of the pockets and assume there are four point-like Fermi
surfaces at $\Gamma$, X, Y and M with finite density of states.  And
we also assume the summation in Eq.~\eqref{gapeq} are only over the
four momentum $\Gamma = (0,0)$, $Y = (\pi, 0)$, $X = (0, \pi)$, and $M
= (\pi, \pi)$.

Then we consider the transformation matrix under this approximation.
In the two-orbital model, the two orbitals, $d_{xz}$ and $d_{yz}$,
mixes strongly in the hole Fermi pockets.  On the other hand, the two
orbitals can be connected by a C$_4$ rotation.  So in the case of
point-like hole Fermi surface, it is obviously that the two orbitals
contribute equally to the hole pockets,
i.e. $U_{h,xz(yz)}[{\Gamma}(M)] = \frac{1}{\sqrt{2}}$, where $h$
denotes the hole band and $xz$ and $yz$ denote the two orbitals.  On
the other hand, the two electron pockets are dominated by d$_{xz}$ and
d$_{yz}$ orbital respectively.  So under the small pocket
approximation, we have $U_{e, xz}(Y) = U_{e, yz}(X) = 1$, and $U_{e,
  yz}(Y) = U_{e, xz}(X) = 0$.  And the nonzero pairing potentials are
$V_{hh}(\Gamma, \Gamma) = V_{hh}(\Gamma, M) = V_{hh}(M, M) =
\frac{v_0}{2}$, $V_{ee}(X, X) = V_{ee}(Y, Y) = v_0$, and
$V_{he}[\Gamma(M), X(Y)] = \frac{v_2}{2}$, where $v_0 = g_0 + 4 g_2$
and $v_2 = g_0 - 4 g_2$.

In the small pocket approximation, the gaps on the two electron
pockets should be same because of the C$_4$ rotational invariance of
the iron pnictide.  Though the gaps on the hole pockets may be
different, we still assume they are equal for simplicity.  So with the
above pairing potentials, we can solve the gap equation \eqref{gapeq}
and get the critical temperature
\begin{align}
  \label{eq:2}
  k_B T_c = 1.14 \omega_D e^{-1/N_h(0)\tilde{v}}
\end{align}
with
$\tilde{v}=\frac{1}{2}[(1+\lambda)(g_0+4g_2)+\sqrt{(1+\lambda)^2(g_0+4g_2)^2-16\lambda
  g_0g_2}]$, where $\lambda = N_e(0)/ N_h(0)$, and $N_e(0)$ and
$N_h(0)$ denote the density of state at the Fermi level of electron
and hole pockets respectively.

\begin{figure}[htbp]
  \center
  \epsfig{figure=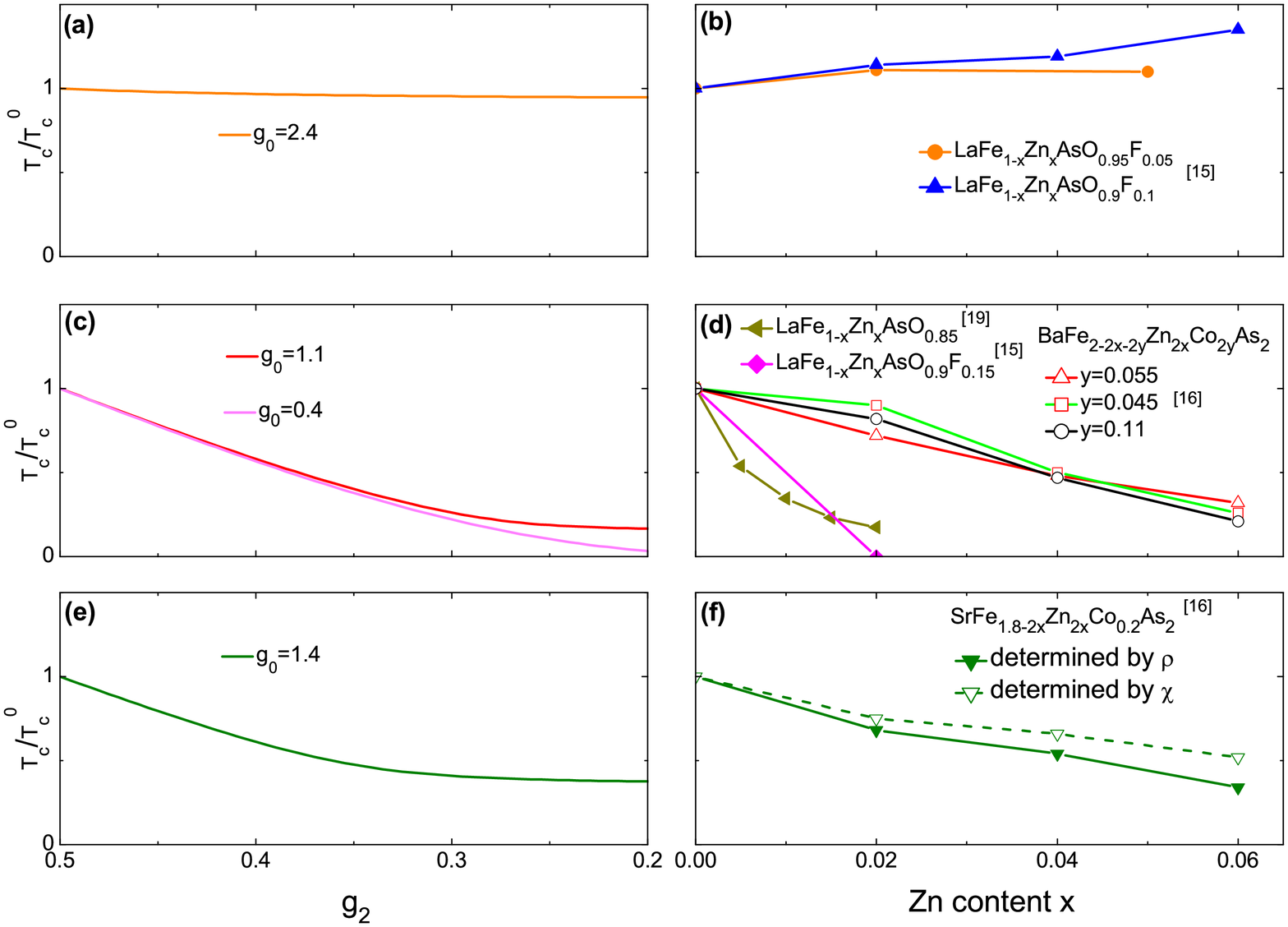,width=0.45\textwidth}
  \caption{\label{fig:tc}(Color online) Left: 	The critical temperature $T_c$ as a function
  of $g_2$ with strong, weak, and moderate on-site pairing coupling $g_0$.
	Right:  Zn-impurity effect on $T_c$ in various Fe-based superconductors observed in experiments. The references of the data are listed.  }
\end{figure}

Our calculation on the critical temperature for various impurity
concentrations, depicted at the left panel of Fig.~\ref{fig:tc}, reveals
that the different impurity-doping behaviors observed in FeSC
~\cite{li:083008,li:671} may be characterized by the strength of the
effective on-site pairing potential $g_0$.  There are three types of
cases for the disorder effect. In the case of large $g_0$, where the
on-site pairing dominates, $T_c$ is hardly suppressed by the Zn
doping. In the case of weak $g_0$, superconductivity is
destroyed by the impurity. When $g_0$ is comparable with $g_2$, as Zn impurity
concentration increases, $T_c$ is initially suppressed rapidly and
then saturate. The experimental facts seem to support the
above scenarios and the effect of Zn doping depends on the material
and the charge carrier concentration . In
LaFe$_{1-x}$Zn$_x$AsO$_{0.9}$F$_{0.1}$ (Ref.~\cite{li:053008}), $T_c$
are insensitive to the Zn-impurity, and may be explained due to large
$g_0$. In the over-doped LaFe$_{1-x}$Zn$_x$AsO$_{0.85}$F$_{0.15}$
(Ref.~\cite{li:083008}) and LaFeAsO$_{0.85}$ (Ref.~\cite{guo:054506}),
in BaFe$_{2(1-x-y)}$Zn$_{2x}$Co$_{2y}$As$_2$ (Ref.~\cite{li:671}),
and in LaFe$_{1-x-y}$Co$_y$Zn$_x$O (Ref.~\cite{yukeliunpublished}),
$T_c$ decreases rapidly with the Zn doping, and may belong to the
category of weak $g_0$. In SrFe$_{1.8-2x}$Zn$_{2x}$Co$_{0.2}$As$_2$
(Ref.~\cite{li:671}), $T_c$ was found to decrease slowly and has the
tendency to saturate although higher Zn-doping will be needed to
confirm the speculation. These scenarios are summarized in
Fig.~\ref{fig:tc}, which shows the critical temperature vs $g_2$ at
different $g_0$ compared with the experimental data of the three types
of materials that behaves differently upon Zn doping.

The moderate value of $g_2$ case is most interesting, for it reflects
the competition between the two SC pairings.  To further explore
this possibility, we have prepared 
SmFe$_{1-x}$Zn$_x$AsO$_{0.9}$F$_{0.1}$ system with $T_C=50K$ and studied
systematically the Zn-impurity effect to $T_c$ experimentally.  The
results are summarized in Fig. 4.  We have measured the change of the
lattice constant due to Zn-doping and confirmed that Zn-atoms are
indeed doped into the iron sites up to $6\%$ of Zn doping, see
Fig. 4(b)~\cite{zn-doping}.  Beyond this doping, our data indicate
that some Zn-impurity may not enter into Fe-lattice so the measurement
of $T_c$ may not correspond to the uniformly doped Zn-impurities.  The
main experimental result of $T_c$ versus Zn concentration on this very
high $T_c$ material is plotted in Fig. 4(a). As we can see, as Zn is
introduced, $T_c$ reduces from 50K continuously down to 40K at $6\%$
of Zn. The slow reduction in $T_c$ may suggest that the
superconductivity saturates at large Zn doping.  It will be
interesting to confirm this by doping high Zn concentration under high
pressure, which remains a challenge in material preparation.

\begin{figure}
  \center
  \epsfig{figure=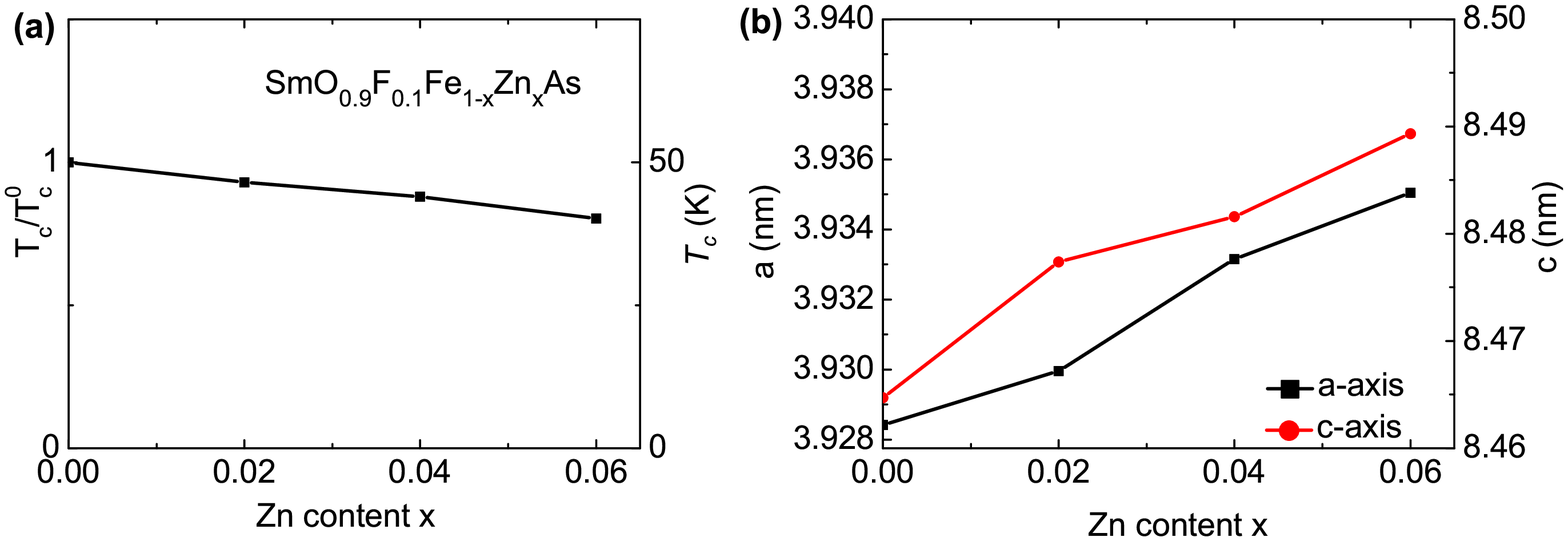,width=0.5\textwidth}
  \caption{(Color online) (a) $T_c$ versus Zn-doping concentration for
    SmFe$_{1-x}$Zn$_x$AsO$_{0.9}$F$_{0.1}$. (b) Lattice constants
    $a$ and $c$ as
    functions of Zn-doping content in
    SmFe$_{1-x}$Zn$_x$AsO$_{0.9}$F$_{0.1}$.}
\end{figure}

\begin{figure}
  \center
  \epsfig{figure=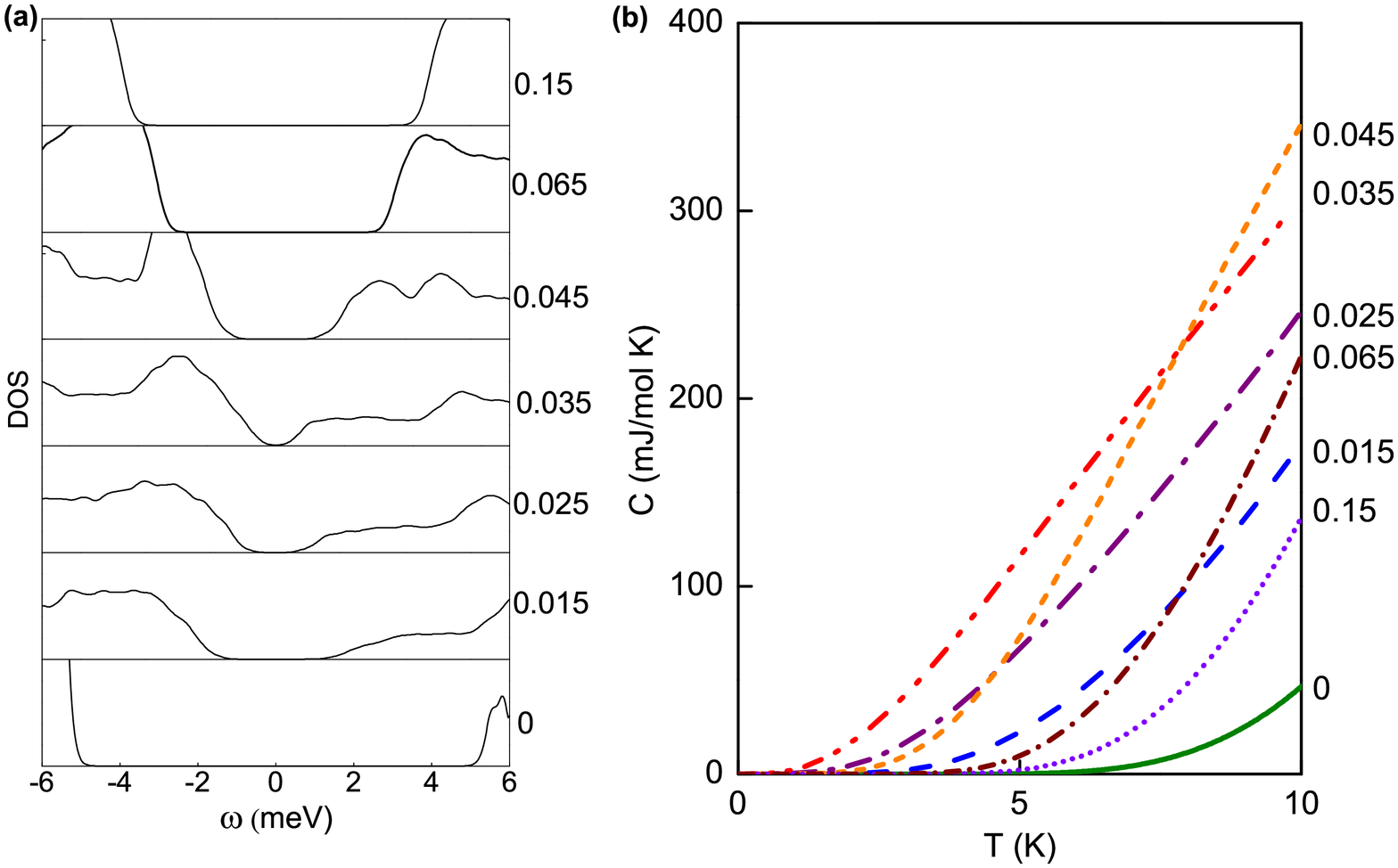,width=0.6\textwidth}
  \caption{\label{fig:dos}(Color online) (a) DOS for various impurity
    concentrations.  $n_{imp} = 0, 0.015, 0.025, 0.035, 0.045, 0.065$ and
    $0.15$ from bottom to top. (b) The low temperature specific heat
    at various impurity concentrations.}
\end{figure}

In addition to the critical temperature, another important feature during
the transition from $s_{\pm}$ to $s_{++}$ is the change of low energy
density of states (DOS).  As shown in Fig.~\ref{fig:dos}(a), the gap
amplitude reduces accompanying with the increase of low energy DOS
when the system approaches the transition point $n_{imp} \approx 0.04$
from clean limit.  And if one further increases the impurity
concentrations, the low energy DOS will be suppressed again due to the
reopening of the gap.  The non monotonic behavior of DOS with
impurities concentration should be able to be observed by integrated
photoemission spectroscopy.

The change of DOS with impurity concentration may also be observed by
the experiments which can measure the low energy DOS, for example the
specific heat.  In the superconducting state, the electron specific
heat can be calculated with
\begin{eqnarray}
\label{eq:6}
	C(T) &=& \frac{\partial}{\partial T} \int_{-\infty}^{\infty} E N(E)f(E)dE,
\end{eqnarray}
where $N(E)$ is the DOS, and $f(E)$ is the Fermi distribution
function.  In the low temperature regime, the temperature dependence
of the superconducting order parameter is very weak and can be
neglected.  So we use the zero-temperature DOS to calculate the
specific heat with Eq.~\eqref{eq:6}.  In the calculation, we use
$\Delta_{\mathrm{coh}} = 0.18 t_1 = 6\mathrm{meV}$ as the energy
scale, and the result is depicted in Fig.~\ref{fig:dos}(b).  We find
that the electron specific heat below $10K$ is small in the clean
limit, and shows a significant increase with approaching the
transition point by increasing impurity concentrations.  When the
system is stabilized in the $s_{++}$ state, $C(T)$ drops to a low
value again.  And the absolute value shown in Fig.~\ref{fig:dos}(b) is
in the same order or even larger than the experimental measurement of
1111 material \cite{wen:174501}.  So it should be able to be observed in
experiments.

We note the recent work of Efremov et al. \cite{efremov:3840}, who
applied T-matrix method to study the non-magnetic effect on FeSC.  Our
microscopic theory shares some similarities with their. In their
phenomenological theory the impurity-doping behavior is found to be
associated with the averaged pairing coupling strength. In our theory,
the decisive role of on-site pairing on the impurity effect is
identified.

\section{summary}
In summary, we have studied the disorder-induced pair-breaking effect
on the Fe-based superconductors using a model that incorporates both
the on-site and NNN pairings. We show that the Zn-impurity largely
suppresses NNN pairing.  Its effect to the superconductivity depends
strongly on the on-site pairing coupling strength $g_0$. The
superconductivity can be robust, or evolves a transition from
$s_{\pm}$ to $s_{++}$, or is strongly suppressed in the presence of
the disorder. Our theory qualitatively explains different reductions
of T$_c$ in various iron pnictide superconductors observed in the
experiments on the Zn-impurity effect.  We also predict the possible
Zn-impurity doping induced transition from $s_{\pm}$ to $s_{++}$
pairing states in certain samples.  Furthermore, we have
systematically prepared Sm-1111 samples with $T_c=50$K under the
Zn-doping and show that the reduction of $T_c$ could be consistent with
the scenario of a moderate on-site s-wave pairing. It will be highly
interesting and important to prepare systematic controlled samples
with higher Zn-doping to experimentally confirm or falsify the theory.
The reduction of the gap in DOS during the transition from $s_{\pm}$
to $s_{++}$ can be observed by integrated photoemission spectroscopy
or specific heat experiments. Finally we remark that the explicit
paring forms are unlikely to be universal in Fe-based superconductors,
in contrast to the universal $d$-wave pairing in cuprates.

\begin{acknowledgements}
This work is partly supported by Hong Kong RGC grant and NSFC/RGC
Joint Research Scheme (No. 10931160425 and No. N$\_$HKU 726/09).
\end{acknowledgements}

\bibliography{ref}

\end{document}